\documentclass[prl,twocolumn,showpacs]{revtex4}

\usepackage{color}
\usepackage{bm}
\usepackage{amsmath}
\usepackage{graphicx}
\usepackage{amssymb}
\usepackage{bm}
\usepackage{times}

\begin{document}

\title{Many-body singlets by dynamic spin polarization}

\author{Wang Yao}
\affiliation{Department of Physics and Center of Theoretical and Computational Physics, The University of Hong Kong, Hong Kong, China}

\begin{abstract}

We show that dynamic spin polarization by collective raising and lowering operators can drive a spin ensemble from arbitrary initial state to many-body singlets, the zero-collective-spin states with large scale entanglement. For an ensemble of $N$ arbitrary spins, both the variance of the collective spin and the number of unentangled spins can be reduced to $O(1)$ (versus the typical value of $O(N)$), and many-body singlets can be occupied with a population of $\sim 20 \%$ independent of the ensemble size. We implement this approach in a mesoscopic ensemble of nuclear spins through dynamic nuclear spin polarization by an electron. The result is of two-fold significance for spin quantum technology: (1) a resource of entanglement for nuclear spin based quantum information processing; (2) a cleaner surrounding and less quantum noise for the electron spin as the environmental spin moments are effectively annihilated.

\end{abstract}

\date{\today}


\maketitle

Many-body singlets (MBS) are the zero-collective-spin states of a spin ensemble with large scale quantum entanglement and zero spin uncertainties. They appear in a variety of contexts in quantum physics and in condensed matter physics, e.g. as horizon states of the quantum black hole~\cite{livine_entanglement_2005}, and as ground states of quantum antiferromagnetic models~\cite{FrustratedSpinSys}.
Their special characteristics place them at the center of attention for quantum applications. \textit{First}, MBS are invariant under a simultaneous unitary rotation on all spins. This makes MBS suitable for spanning a decoherence-free subspace~\cite{bourennane_decoherence-free_2004}, for quantum
communications without a shared reference frame~\cite{bartlett_classical_2003}, and for metrology of the spatial gradient or fluctuations of external fields~\cite{tth_generation_2010}.
\textit{Second}, MBS is an extreme example for the squeeze of spin uncertainties~\cite{wineland_spin_1992,kitagawa_squeezed_1993,sorensen_entanglement_2001,toth_optimal_2007}. The collective spin has zero variance in all directions and thus a source of quantum noise is removed, e.g. in the context of a quantum object affected by a spin bath. \textit{Third}, MBS contain large scale quantum entanglement: every spin is entangled with the rest part of the ensemble. An example of a pure MBS is the product of two-qubit singlets (Bell pairs). In the maximally mixed state of all MBS, the distillable bipartite entanglement is logarithmic in the ensemble size~\cite{livine_entanglement_2005}.

Despite the successful generation of photonic analog of 4-qubit singlets by parametric down conversion~\cite{eibl_experimental_2003,bourennane_decoherence-free_2004}, realization of MBS in a general spin ensemble is an outstanding goal awaiting technically feasible approaches.
Theoretical study shows that spin squeezing based on quantum non-demolition measurement can reduce the total collective spin variance of an atomic ensemble by a factor of 5 in the lossless case~\cite{tth_generation_2010}. However, in such squeezed state the weighting of MBS is small and vanishes in large $N$ limit.

Here we introduce a conceptually new approach for squeezing of collective spin uncertainties and generation of large scale entanglement. The approach uses collective spin raising and lowering operations only, and is applicable to an ensemble of $N$ arbitrary spins initially on arbitrary state. The state after squeezing is significant in figures of merit: in the low loss limit, MBS are occupied with an $N$-independent population of $\sim 20\%$, and both the variance of the total collective spin and the number of spins unentangled with the rest are $O(1)$ (versus the typical values of $O(N)$). We implement this approach in a mesoscopic ensemble of nuclear
spins, a spin system of extensive interests either as a noise source or as a superior information storage in quantum technology. The implementation uses only generic features of dynamic nuclear spin polarization processes by an electron, and is applicable to various electron-nuclear spin systems. Distillation of MBS can be realized by post-selection based on measurement of the electron spin. MBS can be a valuable resource of quantum entanglement for nuclear spin quantum information processing~\cite{Kane_QC_nuclei,Dutt2007_ScienceNV,Newmann08ScienceNVentangle}. In electron spin based quantum computation schemes, preparing the peripheral nuclear spins into MBS results in a cleaner surrounding and hence improved quantum coherence of the electron spin.

We refer to the definition of spin squeezing in the generalized sense~\cite{toth_optimal_2007,tth_generation_2010}, where the degree of squeezing is quantified by $\langle \hat{\bm{J}}^{2} \rangle$, with $\hat{\bm{J}} \equiv \sum_{n=1}^N \hat{\bm{I}}_{n} $ being the total collective spin for an ensemble of $N$ particles with equal or different spins. $\langle \hat{\bm{J}}^{2} \rangle=0$ indicates perfect squeezing where the
$N$ spins are in the MBS. $\langle \hat{\bm{J}}^{2} \rangle \left(\bar{s} \right)^{-1}$ gives an upper bound on the number of spins unentangled with others where $\bar{s}$ is the average spin per particle~\cite{toth_optimal_2007,tth_generation_2010}. States of the spin ensemble can be grouped into multiplets, i.e. irreducible invariant subspaces of the total spin. A multiplet $\{ \left|J,M,\alpha_{J}^k \right\rangle, M= -J, \dots, J \}$ will be denoted in short as $\{ J, \alpha_J^k \}$, where $\alpha_J^k$ is a general index for distinguishing the set of orthogonal (2$J$+1)-dimensional multiplets. The aim is to transfer population from all multiplets to those singlets with $J=0$.

Key to our squeezing approach is to apply raising operator of the form $\hat{j}^+_A-\hat{j}^+_B$ on the spin coherent states $|J, -J, \alpha_J \rangle $. Here the ensemble is partitioned arbitrarily into two subsets $A$ and $B$ with collective spin $\hat{\bm{j}}_{A}$ and $\hat{\bm{j}}_{B}$ respectively (Fig.~1(a)). We find the key identity
\begin{eqnarray}
&& \frac{\left\langle J+1,-J+1,\alpha_{J+1} \right| (\hat{j}^+_A-\hat{j}^+_B )\left|J,-J,\alpha_{J} \right\rangle }{ \left\langle J,-J,\alpha_{J} \right| (\hat{j}^+_A-\hat{j}^+_B )\left|J+1,-J-1,\alpha_{J+1} \right\rangle ^{*}} \nonumber \\
&& =  -\left [ \left(J+1\right)\left(2J+1\right) \right ]^{-\frac{1}{2}}. \label{eq:ratio2}
\end{eqnarray}
Moreover, $\left\langle J^{\prime},M^{\prime},\alpha_{J'} \right|\hat{j}^+_A-\hat{j}^+_B\left|J,-J,\alpha_{J}\right\rangle =0$
for $|J^{\prime} - J| > 1$ or $M' \neq -J+1$. Thus, under the condition that each multiplet is initialized on the spin coherent state, application of the $\hat{j}^+_A-\hat{j}^+_B$ operator tends to transfer populations from multiplets of larger dimension to multiplets of smaller dimension (Fig.~1(c)). The transfer rate of $\{J+1,\alpha_{J+1}^q\} \rightarrow \{J,\alpha_J^{k}\}$ is by a factor of $(J+1)(2J+1)$ larger than that of the backward transfer $\{J,\alpha_J^k\} \rightarrow \{J+1,\alpha_{J+1}^{q}\}$.

\begin{figure}[t]
\includegraphics[bb=85bp 180bp 505bp 615bp,clip,width=8cm]{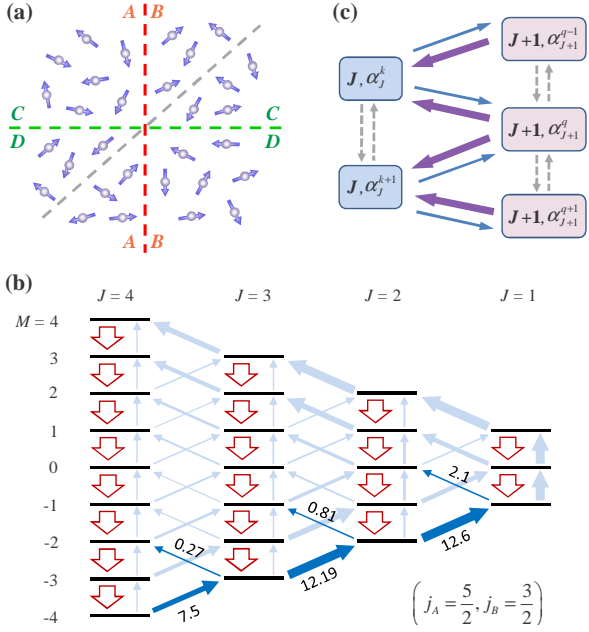}
\caption{(a) The red dashed line partitions the spin ensemble into two subsets $A$ and $B$ with collective spin $\hat{\bm{j}}_{A}$ and $\hat{\bm{j}}_{B}$ respectively. The green dashed line gives a different partition of the ensemble into subset $C$ and $D$. (b) An example of how the operator $\hat{j}^+_A-\hat{j}^+_B$
couples various basis states $\left|J,M,j_A,j_B\right\rangle $
(solid arrows). The absolute value squared of the transition matrix elements are indicated with the thickness of the arrows. The transitions related to the spin coherent states $\left|J,-J\right\rangle $ are
highlighted. Hollow vertical arrows show the coupling by $\hat{J}^{-}$.
(c) Schematics of the population transfer rates between multiplets under the condition that each multiplet is initialized on the spin coherent state $|J, -J \rangle$ when the $\hat{j}^+_A-\hat{j}^+_B$ operator is applied. The transfer rate of $\{J,\alpha_J^k\} \rightarrow \{J,\alpha_J^{k'}\}$ is identical to the rate of the backward transfer $\{J,\alpha_J^{k'}\} \rightarrow \{J,\alpha_J^{k}\}$. The rate of $\{J+1,\alpha_{J+1}^q\} \rightarrow \{J,\alpha_J^{k}\}$ is by a factor of $(J+1)(2J+1)$ larger than that of the backward transfer $\{J,\alpha_J^k\} \rightarrow \{J+1,\alpha_{J+1}^{q}\}$.}
\label{illustration}
\end{figure}

Squeezing of collective spin uncertainties can therefore be realized by dynamic spin polarization with the lowering operator $\hat{J}^{-}$ and raising operators of the form $\hat{j}^+_A-\hat{j}^+_B$. Consider the use of two such operators $\hat{j}^+_A-\hat{j}^+_B$ and $\hat{j}^+_C-\hat{j}^+_D$ where $C$ and $D$ constitute a different bipartition of the ensemble (Fig.~1(a)). The Hilbert space can be divided into independent subspaces according to the quantum numbers $\{j_{A\cap D},j_{B\cap D},j_{B\cap C},j_{A\cap C}\}$ conserved by the raising/lowering operations, and their values determine the number of (2$J$+1)-dimensional multiplets $n(J)$. If $\hat{J}^{-}$ is applied more frequently such that the system is in spin coherent states every time $\hat{j}^+_A-\hat{j}^+_B$ or $\hat{j}^+_C-\hat{j}^+_D$ is applied, we find the steady-state in each subspace: $\rho= \sum_{J} f(J) \sum_{k=1}^{n(J)} |J,-J,\alpha_J^k \rangle \langle J,-J,\alpha_J^k | $ where $f(J)= (J+1)(2J+1) f(J+1)$.
Most subspaces contain at least one MBS~\cite{remark}, and $n\left(J\right) \leq n\left(0 \right)\left(2J+1\right)$. Thus, in the steady state, MBS are occupied with a population $ n(0) f(0) \geq \left[\sum_{J} g (J) \right]^{-1}=0.20 $, and the variance $ \langle \hat{\bm{J}}^{2} \rangle \leq\left[\sum_{J} g(J) \right]^{-1}\sum_{J} J\left(J+1\right) g(J)=2.44$,
where $g (J) \equiv (2J+1)\left[\prod_{i=0}^{J-1}\left(i+1\right)\left(2i+1\right) \right]^{-1}$.

\begin{figure}[t]
\includegraphics[bb=48bp 203bp 555bp 626bp,clip,width=8cm]{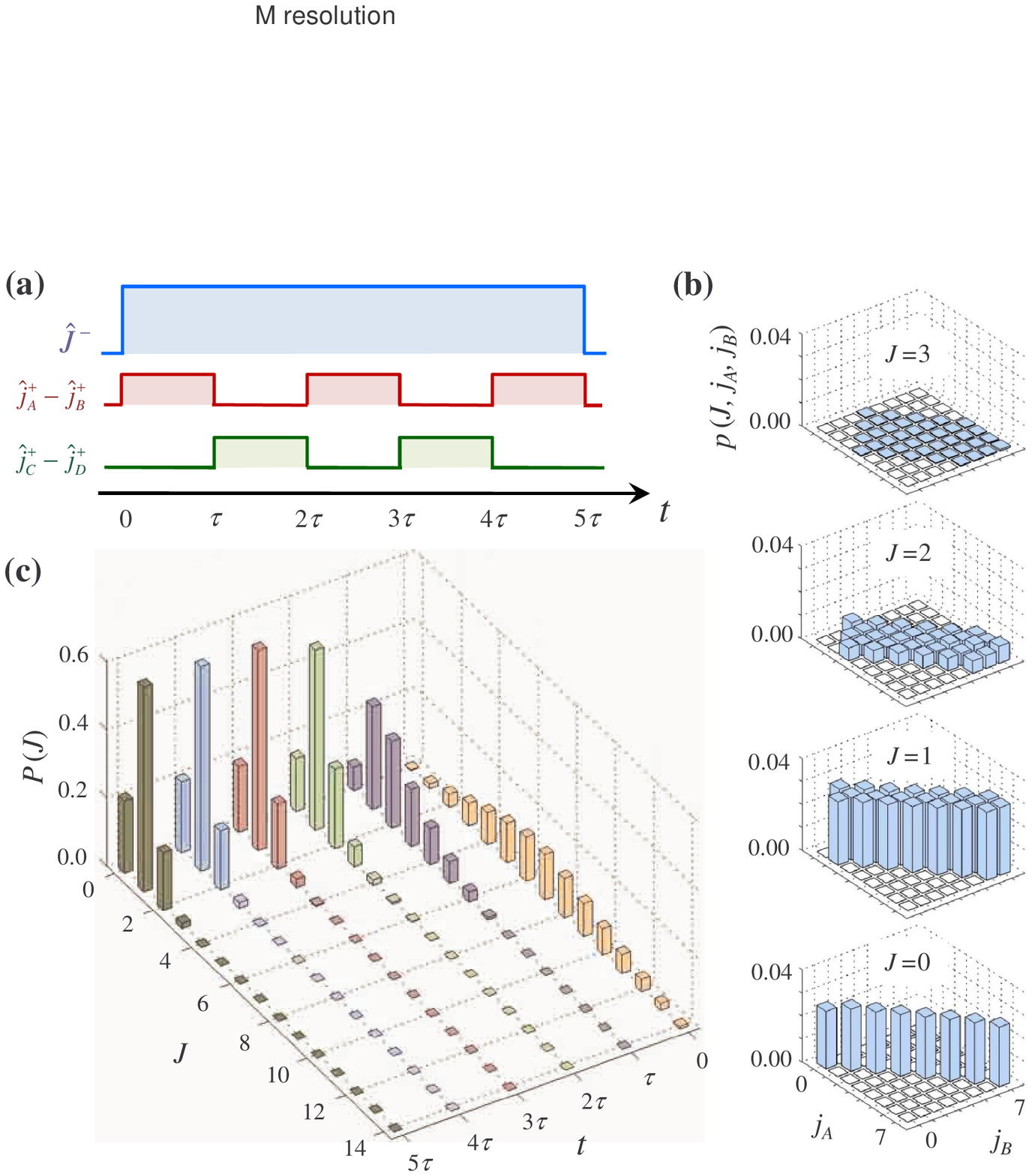}
\caption{(a) An example of the squeezing control by dynamic spin polarization with the operators $\hat{J}^-$, $\hat{j}^+_A-\hat{j}^+_B$ and $\hat{j}^+_C-\hat{j}^+_D$. (b) and (c) show simulation results of the control in (a). The system is initially in the completely mixed state in the subspace with $\{j_{A\cap D}=7/2, j_{B\cap D}=7/2, j_{B\cap C}=7/2, j_{A\cap C}=7/2 \}$. The polarization rates: $10^{-3}\Lambda_h=\Lambda_o$, and the delay
time $\tau = 2/\Lambda_o$. (b) $p( J,j_A,j_B )$ gives the population of the multiplet $\{ J, j_A, j_B\}$ in the state at $t=4\tau$. (c) $P(J) \equiv \sum_{j_A,j_B}p(J,j_A,j_B)$ gives the integrated probability of finding the system with total collective spin $J$ at various time. At $t=5\tau$, MBS are occupied with the population $P(J=0) = 0.21$, and $\langle \hat{\bm{J}}^{2} \rangle = 2.42$. }
\label{simulation}
\end{figure}

Dynamic spin polarization by the collective raising and lowering operators can be described by Lindblad terms in the master equation $\dot{\rho} = -\frac{1}{2}\sum_{m=1}^{3} (\hat{L}_{m}^{\dagger}\hat{L}_{m}\rho+\rho\hat{L}_{m}^{\dagger}\hat{L}_{m}-2\hat{L}_{m}\rho\hat{L}_{m}^{\dagger} )$ where $\hat{L}_{1}\equiv\sqrt{\Lambda_h}\hat{J}^{-}$, $\hat{L}_{2}\equiv \sqrt{\Lambda_o} (\hat{j}^+_A-\hat{j}^+_B )$,
and $\hat{L}_{3}\equiv \sqrt{\Lambda_o} (\hat{j}^+_C-\hat{j}^+_D )$. The $\hat{J}^-$ and $\hat{j}^+_A -  \hat{j}^+_B$ operators are applied with the rates $\Lambda_h |\langle \psi_f |\hat{J}^- | \psi_i \rangle |^2 $ and $ \Lambda_o |\langle \psi_f |\hat{j}^+_A -  \hat{j}^+_B| \psi_i \rangle |^2$ respectively, and the squeezing scheme requires that the former rate shall always be larger. We note that
$\langle J, M, j_A, j_B | (\hat{j}_A^- - \hat{j}_B^-)(\hat{j}^+_A - \hat{j}^+_B) | J, M, j_A, j_B \rangle$ increases with the decrease of $J$ and reaches the maximal value of $\sim (j_A + j_B)^2$ for small $J$, while $\langle J, M |\hat{J}^+  \hat{J}^- | J, M \rangle \sim J^2$. Thus we find the requirement $\Lambda_h/\Lambda_o > (j_{A\cap D} + j_{B\cap D} + j_{B\cap C} + j_{A\cap C})^2 $, the latter quantity $\sim 4N s^2$ in an ensemble of $N$ spin $s$ particles. Spin decoherence causes the population decay of MBS with a rate $\sim N \gamma_n$ with $\gamma_n$ being the single spin decoherence rate. The low loss condition is therefore defined as $\frac{1}{4N s^2}\Lambda_h > \Lambda_o \gg \gamma_n$, where spin decoherence has negligible effect on the squeezing efficiency~\cite{Yu_inprep}.

We numerically demonstrate a squeeze control where $\hat{j}^+_A-\hat{j}^+_B$ and $\hat{j}^+_C-\hat{j}^+_D$ are applied in alternating fashion (see Fig.~2(a)). With spin decoherence neglected under the low loss condition, calculation can be significantly simplified for this choice of control and a moderately large spin system can be simulated. In the interval when $\hat{J}^-$ and $\hat{j}_A^+ - \hat{j}_B^+$ (or $\hat{j}_C^+ - \hat{j}_D^+$) are applied, the relevant Hilbert space can be further divided into independent subspaces according to the quantum numbers $\{j_A, j_B \}$ (or $\{j_C, j_D\}$). If the duration $\tau$ of each interval is sufficiently large to ensure the reach of steady state, we simply need to solve for steady state in each small subspace and keep track of the basis transform between $\{|J,M, j_A, j_B \rangle \}$ and $\{|J,M, j_C, j_D \rangle \}$ upon the switch of raising operators. Off-diagonal coherence is found to be negligible which can further simplify the calculation. An example of the simulated squeezing dynamics is given in Fig. 2(b-c). The initial density matrix is the completely mixed one in the subspace with $\{j_{A\cap D}=7/2, j_{B\cap D}=7/2, j_{B\cap C}=7/2, j_{A\cap C}=7/2 \}$. The polarization rates are $10^{-3}\Lambda_h=\Lambda_o$, and $\tau = 2/\Lambda_o$. Fig.~2(b) shows that the population distribution $p(J,j_A,j_B)$ among the multiplets indeed approaches the steady state value after a time of $4\tau$. At $t=5\tau$, MBS are occupied with a population of $0.21$ and $\langle \hat{\bm{J}}^{2} \rangle = 2.42$.

In an ensemble of nuclear spins, the applications of the two types of operators $\hat{J}^{-}$ and $\hat{j}_{A}^+-\hat{j}_{B}^+$ are realized in the process of dynamic nuclear spin polarization (DNSP), a major tool for manipulation of nuclear spins~\cite{greilich_nuclei-induced_2007,reilly_suppressing_2008,xu_optically_2009,tartakovskii_nuclear_2007,danon_nuclear_2008,korenev_nuclear_2007,bracker_optical_2005,urbaszek_efficient_2007,Ono_current_oscillation,latta_confluence_2009,vink_locking_2009,laird_hyperfine-mediated_2007,rudner_electrically_2007,rashba_theory_2008}. We consider the hyperfine interaction $\hat{H}_{0}=\sum_{n}\left|\psi\left(\bm{r}_{n}\right)\right|^{2}\hat{\bm{I}}_{n}\cdot\overleftrightarrow{\bm{A}}\cdot\hat{\bm{S}}$ coupling the electron spin $\hat{\bm{S}}$ to peripheral lattice nuclear spins $\hat{\bm{I}}_{n}$. $\overleftrightarrow{\bm{A}}$
is the hyperfine coupling constant in tensor form, and the position
dependence of coupling enters through the envelope function $\psi\left(\bm{r}\right)$
of the electron only. $\hat{H}_{0}$ describes generally the hyperfine interaction of electron or hole system in quantum dots or shallow donors formed in group IV or III-V materials~\cite{Chamarro_holespin,fischer_spin_2008}. In most DNSP schemes, $\hat{H}_{0}$ induces the electron-nuclear flip-flop in passing electron spin polarization to the nuclei and the energy cost is compensated by emission/absorption of phonons or photons~\cite{tartakovskii_nuclear_2007,danon_nuclear_2008,korenev_nuclear_2007,bracker_optical_2005,urbaszek_efficient_2007}.
These DNSP schemes are termed as the \textit{dc} type hereafter. Alternatively, DNSP can also utilize the \textit{ac} correction to the hyperfine coupling: $\hat{H}_{ac}=\sum_{n}\left(\bm{d}_{\omega}\cdot\nabla\left|\psi\left(\bm{r}_{n}\right)\right|^{2}\right)\cos(\omega t)\hat{\bm{I}}_{n}\cdot\overleftrightarrow{\bm{A}}\cdot\hat{\bm{S}}$ when an \textit{ac} electric field
induces an electron displacement $\bm{d}_{\omega}\cos(\omega t)$, with energy cost for electron-nuclear flip-flop directly supplied by \textit{ac} field~\cite{laird_hyperfine-mediated_2007,rudner_electrically_2007,rashba_theory_2008}. Such DNSP process is termed hereafter as the \textit{ac} type.

For nuclear spins on the periphery of an electron, MBS can be realized by combining \textit{dc} and \textit{ac} DNSP processes which polarize nuclear spins in opposite directions with the operators $\sum_{n}\left|\psi\left(\bm{r}_{n}\right)\right|^{2}\hat{I}_{n}^{-}$ and $\sum_{n} \frac{\partial}{\partial\mu}\left|\psi\left(\bm{r}_{n}\right)\right|^{2} \hat{I}_{n}^{+}$ respectively. Here $\mu$ is the direction of \textit{ac} electric field. The lattice sites with equal electron density $\left|\psi\left(\bm{r}\right)\right|^{2}$
are grouped into coordination shells. On each shell, $\sum_{n}\left|\psi\left(\bm{r}_{n}\right)\right|^{2}\hat{I}_{n}^{-}$
and $\sum_{n} \frac{\partial}{\partial\mu}\left|\psi\left(\bm{r}_{n}\right)\right|^{2} \hat{I}_{n}^{+}$
are of the character of $\hat{J}^{-}$ and $\hat{j}_{A}^{+}-\hat{j}_{B}^{+}$ respectively.
Under the influence of incoherent electron spin dynamics in the DNSP process, the large shell-to-shell difference in the \textit{dc} hyperfine coupling strength causes loss of inter-shell coherence in a timescale much faster than the squeezing. Thus different coordination shells can be independently squeezed towards MBS.

\begin{figure}[t]
\includegraphics[bb=46bp 446bp 555bp 779bp,clip,width=8cm]{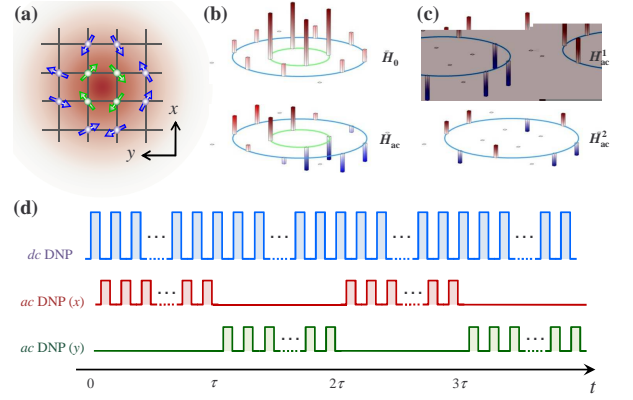}
\caption{(a) Schematics of an electron in a nuclear spin bath.
The first coordination shell has 4 nuclear spins (green color) and
the second shell has 8 nuclear spins (blue color). (b) Upper part: \textit{dc}
hyperfine coupling coefficients at the various lattice sites. Lower
part: \textit{ac} hyperfine coupling coefficients with \textit{ac}
electric field in $x$ direction. The heights of the bar give the
magnitude. (c) The \textit{ac} hyperfine coupling
decomposed into two terms $\hat{H}_{ac}=\hat{H}_{ac}^{1}+\hat{H}_{ac}^{2}$. $\hat{H}_{ac}^{1}=\tilde{a}_1(\hat{j}^+_A-\hat{j}^+_B)\hat{S}^{-}e^{i\omega t}E_{x}+c.c.$, and
$\hat{H}_{ac}^{2}=\tilde{a}_2(\hat{j}_{C}^{+}-\hat{j}_{D}^{+})\hat{S}^{-}e^{i\omega t}E_{x}+c.c.$.
The 4 sites with positive coupling coefficients in the upper part form subset $A$ and the 4 sites with positive
coupling in the lower part form subset $C$, and $B$ ($D$) is the complement of $A$ ($C$).
(d) Schematic of the DNSP control where the switching between \textit{dc}
and \textit{ac} DNSP is concatenated with the switching
of the \textit{ac} electric field between $x$ and
$y$ directions.}
\label{nuclearspin}
\end{figure}

Fig.~3(a) shows the schematic of an electron with a 2D Gaussian envelope
function. The 12 lattice nuclear spins form two coordination shells
according to the \textit{dc} hyperfine coupling strength (Fig.~3(b)).
For the green shell with 4 lattice nuclear spins, $\hat{H}_{ac}=\tilde{a}\hat{S}^{-}(\hat{j}^+_A-\hat{j}^+_B)e^{i\omega t}E_{x}+\tilde{a}\hat{S}^{-}(\hat{j}^+_C-\hat{j}^+_D)e^{i\omega t}E_{y}+c.c.$,
where $E_{x(y)}$ is the \textit{ac} electric field
in the $x$ ($y$) direction. Fig.~3(d) shows the schematic of the DNSP control
where the switching between \textit{dc} and \textit{ac} DNSP is concatenated with the switching of the \textit{ac} electric field between $x$ and $y$ directions. Numerical simulation of such a control has been given in Fig.~2.
The blue shell represents the more general case where nuclear spins are polarized in the \textit{ac} DNSP process by $\tilde{a}_1(\hat{j}^+_A-\hat{j}^+_B)+\tilde{a}_2(\hat{j}_{C}^{+}-\hat{j}_{D}^{+})$, a linear superposition of raising operators of the desired form (Fig.~3(c)). The identity in Eq. (\ref{eq:ratio2}) obviously holds if $\hat{j}^+_A-\hat{j}^+_B$
is replaced by $\hat{j}_{C}^{+}-\hat{j}_{D}^{+}$, thus we have this same identity for their linear superposition as well. Numerical simulations confirm that operators of this new form are equally efficient in the squeezing effect~\cite{Yu_inprep}.

Interaction between neighboring nuclear spins causes spin diffusion and spin dephasing which can result in loss of MBS. The dipolar interaction between neighboring lattice sites is of the strength $\sim 10$ Hz. Nuclear spin diffusion by the $\hat{I}_n^+ \hat{I}_m^-$ coupling terms is efficiently suppressed when the shell-to-shell inhomogeneity in the hyperfine coupling is large. By the $\hat{I}_n^z \hat{I}_m^z$ term, the nuclear spins are subject to a dipolar magnetic field dependent on the configuration of their neighbors, which leads to dephasing with a rate $\gamma_n \sim 10-100$ Hz. To realize efficient squeezing, fast DNSP mechanisms are desired.

For optically controllable electron spin, e.g. in quantum dot or impurity in III-V semiconductors, fast \textit{dc} DNSP can be realized by the hyperfine-mediated optical excitation of spin-forbidden excitonic transitions~\cite{korenev_nuclear_2007,chekhovich_pumping_2010}. Assuming the electron Zeeman splitting $\omega_e \sim 0.2 $ GHz, the intrinsic broadening of charged exciton $\gamma_t \sim 0.2 $ GHz, and an optical Rabi frequency $\Omega\sim 3 $ GHz for the excitonic transition, we estimate the DNSP rate: $\Lambda_h=\frac{a^2 \Omega^2}{\omega_e^2 \gamma_t} \sim 10$ MHz on a coordination shell with hyperfine coupling $a = 3$ MHz. For other electron-nuclear spin system, fast \textit{dc} DNSP may be realized through the bath-assisted electron-nuclear flip-flop in the presence of an efficient energy dissipation channel, e.g. an electron Fermi sea in nearby leads~\cite{danon_nuclear_2008}.

\textit{ac} DNSP is of the rate $\Lambda_o=\frac{\tilde{a}^2}{\gamma_s}$ where $\gamma_s$ is the broadening of the electron spin resonance~\cite{rudner_electrically_2007}. The magnitude of the \textit{ac} hyperfine interaction, $\tilde{a}$, depends on the strength of \textit{ac} electric field and the inhomogeneity of the electron envelop function. Giving the phosphorus donor in silicon as an example, we have the first several shells: $(A,6.0, 6)$, $(B,4.5,12)$, $(C,3.3,4)$, $(D,2.2,12)$ and $(F,1.7,12)$ where the first letter is the label of the shell by convention, the second number is the hyperfine coupling strength in unit of MHz, and the third is the number of equivalent sites on the shell~\cite{hale_shallow_1969}. The distance between neighboring shells is in the order of $\sim 0.1$ nm. Thus, we estimate $\tilde{a} \sim $ MHz by a moderate displacement of the electron $d_{\omega} \sim 0.1$ nm. Assuming $\gamma_s \sim 0.1$ GHz, $\Lambda_o$ can be of $\sim 10$ kHz on these shells. Thus, we conclude that the low loss condition can indeed be satisfied for nuclear spins on the periphery of a strongly confined electron.

The work was supported by the Research Grant Council of Hong Kong under Grant No. HKU 706309P. The author acknowledges helpful discussions with L. J. Sham, H. Y. Yu and X. D. Xu.


\begin{thebibliography}{31}
\bibitem{remark} The states unconnected with MBS are of $\sim 1 \%$ of the entire Hilbert space if two raising operators of the form $\hat{j}_{A}^{+}-\hat{j}_{B}^{+}$ are used, and can be reduced to $\sim 0.01 \%$ if three such operators are used in the dynamic spin polarization.
\end{thebibliography}
\end{document}